\documentclass[letter]{aa} 
\usepackage{graphicx}
\usepackage{txfonts}
\begin{document}
   \title{White dwarf masses derived from planetary nebulae modelling}

   \titlerunning{White dwarf masses from planetary nebulae}

   \author{K. Gesicki
          \inst{1}
          \and
          A.A. Zijlstra\inst{2}
          }


   \institute{Centrum Astronomii UMK,
           ul.\,Gagarina 11, PL-87-100\, Torun, Poland \\
              \email{Krzysztof.Gesicki@astri.uni.torun.pl}
         \and
             School of Physics and Astronomy, University of Manchester,
             P.O. Box 88, Manchester M60 1QD, UK\\
             \email{a.zijlstra@manchester.ac.uk}
             }

   \date{Received ; accepted}

 
  \abstract 
{} 
{We compare the mass distribution of central stars
  of planetary nebulae (CSPN) with those of their progeny, white dwarfs
  (WD). } 
{We use a dynamical method to measure
  masses with an uncertainty of 0.02\,M$_\odot$. } 
{The CSPN mass distribution is sharply peaked at $0.61\,\rm
  M_\odot$. The WD distribution peaks at lower masses ($0.58\,\rm M_\odot$) and
  shows a much broader range of masses.  Some of the difference can be
  explained if the early post-AGB evolution is faster than predicted by the
  Bl\"ocker tracks.  Between 30 and 50 per cent of WD may avoid the PN phase
  because of too low mass. However, the discrepancy cannot be fully resolved
  and WD mass distributions may have been broadened by observational or model
  uncertainties.}  
{}


   \maketitle
%

\section{Introduction}

White Dwarf (WD) mass distributions have been determined using a variety
of different methods. Discrepancies exist between the different
determinations in particular between the photometric and spectroscopic
WD masses. Boudreault \& Bergeron (2005) compared the masses derived by
fitting the observed Balmer lines with masses derived from trigonometric
parallaxes and photometry. They found differences of $\sim 50$ per cent
for cool (6\,500--14\,000\,K) DA white dwarfs. Spectroscopic masses
are believed to be more accurate, especially for WDs in the temperature
range between 15\,000 and 40\,000\,K (Liebert et al. 2005). Atmospheric
models are less well established for stars outside this range.  For
hotter WDs the atmospheric structure is modified by an (often unknown)
amount of metals and by non-LTE effects. For cooler WDs the convection
has to be considered and the models are sensitive to the mixing length
and the amount of helium convected to the surface (Boudreault \&
Bergeron 2005).

Central stars of planetary nebulae (CSPN) provide a way to test the mass
distributions.  CSPNe evolve directly into WDs, with only very minor
mass changes, allowing one to measure masses of currently forming white
dwarfs. However, CSPN mass distributions have also been uncertain. For
example, Napiwotzki (2006) shows that the very high CSPN masses (close
to the Chandrasekhar limit) derived spectroscopically with
state-of-the-art model atmospheres by Pauldrach et al. (2004) are
physically implausible and masses close to the peak of the CSPN/WD
mass distribution are more likely.

CSPN masses are normally obtained from the luminosities.  But more accurate
masses can be derived using the age--temperature diagram, obtainable from the
surrounding planetary nebula (PN). Gesicki et al. (2006) applied this to a
sample of 101 PNe.  In this Letter we discuss the resulting mass distributions
for hydrogen-rich and hydrogen-poor CSPNe and compare with published WD
masses.

\section{Methods and results}

\subsection{Models}

The method requires the age of the nebula and the temperature of the
central star to be determined. Together these provide the heating
time scale for the star.

We derive the age of the PNe using a combination of line ratios,
diameters (taken from the literature), and new high resolution spectra
(Gesicki et al. 2006). The diameters and line ratios are used to
fit a spherically symmetric photo-ionization model. The model assumes a
density distribution and finds a stellar black-body temperature. For each ion,
the model finds a radial emissivity distribution.  The observed line profiles
for each ion represent the convolution of the thermal broadening and
the expansion velocity at each radius. Thus, the line profiles for different
ions are used to fit a velocity field.  An iterative procedure is used to
improve the ionization model. The emissivity distributions of different ions
overlap, and this gives a strong constraint on the shape of the wings of the
line profiles. A genetic algorithm, PIKAIA, is used to arrive at the optimum
solution for ionization model and velocity field. A turbulent component is
added if needed: turbulence is indicated by a Gaussian shape of the line
profiles. The expansion velocities are found to increase with radius, due to
the overpressure of the ionized region.

>From the velocity field $v(r)$, we derive the mass-weighted average over the
nebula, $v_{\rm av}$. This parameter has been shown to be robust against the
simplifications. Different models which provide comparable quality fits give
the same $v_{\rm av}$ to within 2\,km\,s$^{-1}$ (Gesicki et al 2006).
Applying this to a radius of 0.8 times the outer radius (equivalent to the
mass-averaged radius) allows us to define a kinematic age $t$ to the nebula.
A linear acceleration is assumed to have occurred from the AGB expansion
velocity (10--15\,km\,s$^{-1}$) to the PN velocity $v_{\rm av}$
(20--25\,km\,s$^{-1}$).

The derived nebular  age and stellar temperature are compared to the the
H-burning tracks of Bl\"ocker (1995), which provide the largest and most
uniform collection available. We interpolate between different tracks to
find for each ($t$,$T_{\rm eff}$), the CSPN luminosity and mass. 

\subsection{Different CSPN types}

The CSPNe fall into two broad categories: the hydrogen-rich O-type stars
and the emission-line central stars which are generally
hydrogen-deficient. The second group consists of [WR]-type stars with
strong emission lines and {\it wels} (weak emission line stars). The
[WR] are subdivided into hot [WO] and cool [WC]. [WR] stars are in most
cases hydrogen-free (three possible exceptions are mentioned by Werner
\& Herwig 2006). The {\it wels} may contain some hydrogen. Gesicki et
al. (2006) show that one group of {\it wels} is located in the
temperature gap between [WC] and [WO] stars. The other {\it wels} stars
form a non-uniform group, including higher-mass objects where the high
luminosity drives a wind but the star is not necessarily hydrogen-poor.
The hydrogen-rich stars are believed to be related to the DA white
dwarfs, while the [WR] may evolved into DB's.

\subsection{The HR diagram}

The full analyzed sample contains 101 PNe, of which about 60 are in the
direction of the Galactic Bulge and the remainder are in the Galactic
disk. Foreground confusion among the Bulge PNe is estimated at 20\%. The
sample contains 23 [WR]-type, 21 {\it wels} and 57 non-emission-line central
stars\footnote{ The data file is available from web page {\tt
www.astri.uni.torun.pl/$\sim$gesicki/modelled$\_$pne.dat}}. The CSPN
classification was adopted from literature. The last group contains also
objects without any information about their spectrum.

   \begin{figure}
   \includegraphics[width=88mm]{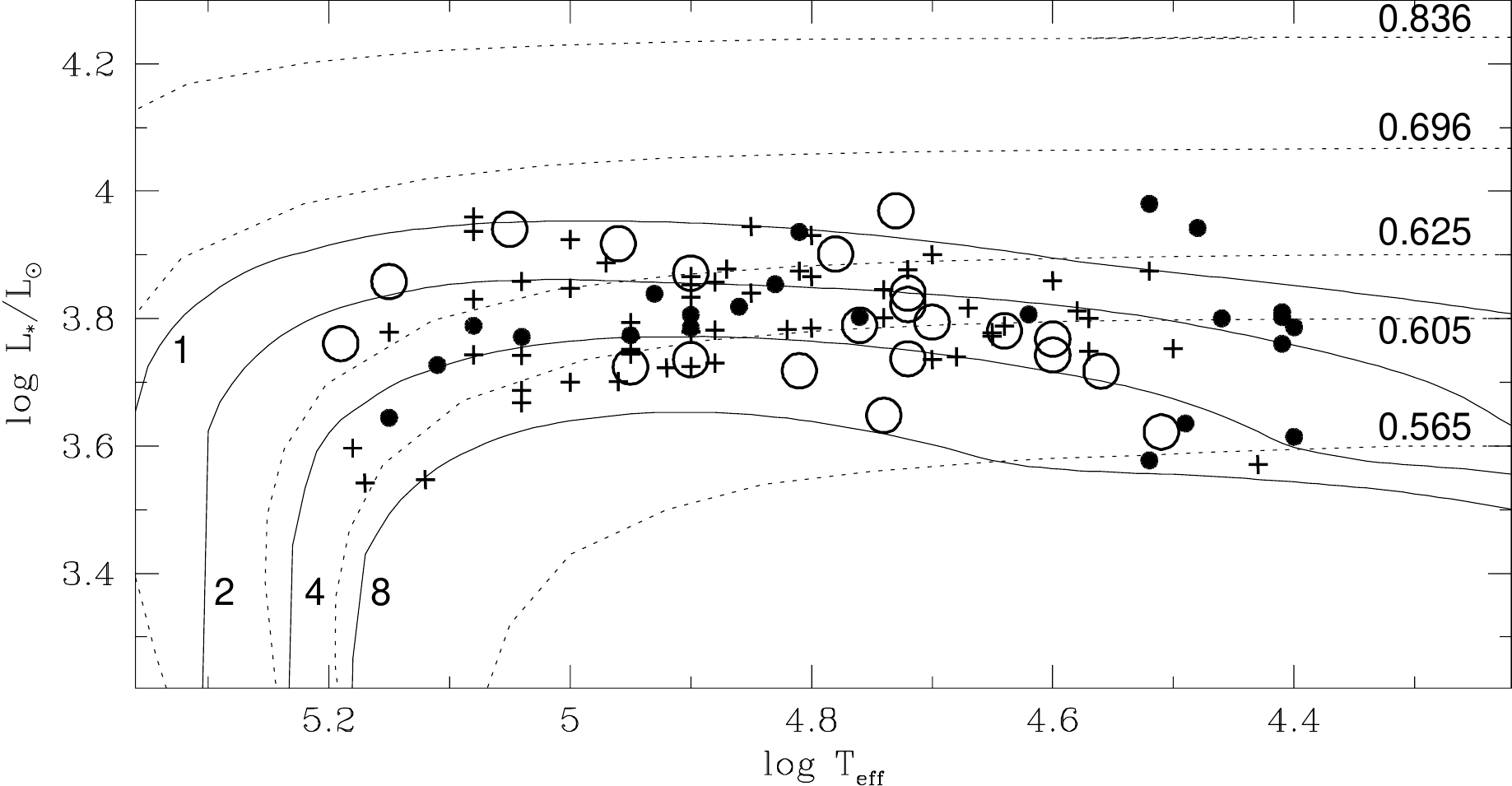}
   
     \caption{Comparison of the 101 modelled PNe with the evolutionary
     tracks in the HR diagram. The model black-body temperatures
     are plotted against the luminosities interpolated from tracks.
     Filled circles indicate [WR] stars, open circles are {\it wels} and
     pluses indicate non-emission-line stars. The dotted lines show 
     H-burning evolutionary models of Bl\"ocker (1995), labeled by mass
     in units of M$_\odot$. The solid lines are isochrones, labeled by
     the time after the nebula ejection, in units of $10^3$\,yr.}

         \label{hr}
    \end{figure}

 In Fig.\ref{hr} we show the photoionization temperatures and
interpolated luminosities, plotted on the HR diagram. The H-burning
tracks of Bl\"ocker (1995) are also shown: the luminosities and masses
of CSPNe fall into a rather restricted range of values. Isochrones of
1,2,4, and $8\times 10^3$\,yr are also shown. 

A previous HR diagram of CSPNe presented by Stanghellini et al. (2002)
shows a much broader range of luminosities and, in consequence, masses.
They use Zanstra temperatures and luminosities.  The Zanstra method of
locating a CSPN in the HR diagram was criticized by Sch\"onberner \&
Tylenda (1990). Observationally, the accuracy of the luminosity
determinations is about a factor of 2. On the Sch\"onberner tracks, a
CSPN mass change from, e.g., 0.57 to 0.7\,M$_\odot$ corresponds to a
factor of 3 in luminosity. The masses determined directly from
luminosity are thus accurate to only 0.1\,M$_\odot$. This is less than
the typical dispersion of masses.  In contrast, for the same mass range,
the dynamical time scales differ by a factor of 60. Even for a factor of
2 uncertainty in the nebular age, the mass changes by only
0.02\,M$_\odot$. Therefore, the dynamical method improves the accuracy.

Sch\"onberner \& Tylenda (1990) also developed a method to improve the
CSPN mass determination. This method (Tylenda et al. 1991) results in
masses similar to ours.

Table \ref{sp_mass} compares, for four objects in common, our dynamical
masses with the spectroscopic masses derived by Kudritzki et al. (2006).
The spectroscopic masses are larger, in two cases very much larger.
The lower masses are supported by the kinematical properties of 
Tc\,1 and He\,2-108 (see Fig.\,5 of Napiwotzki 2006), which favour 
an old thin disk population. Kudritzki et al. also derive $T_{\rm eff}$: our
photo-ionization values are in good agreement.

Pauldrach et al. (2004) find from a spectroscopic analysis, five CSPNe
with masses close to the Chandrasekhar limit. This result is
implausible, as argued by Napiwotzki (2006).  Three of their objects are
also in our sample, and all are found to have regular masses.

\begin{table}
\caption[]{\label{sp_mass} Comparison between our dynamical masses
and spectroscopic masses from Kudritzki et al. (2006). Observed
mass-loss rates from the same paper are also listed and compared to
values from the model tracks of Bl\"ocker (1995). He\,2-108 is classified 
as {\it wels}, the other three are non-emission-line stars.
  }
\begin{flushleft}
\begin{tabular}{llllllllll}
\hline
\noalign{\smallskip}
 Object & \multicolumn{2}{c}{$M$\,[M$_\odot$]} & 
  \multicolumn{2}{c}{$T_{\rm eff}$\,[$10^3$\,K]} &  
  \multicolumn{2}{c}{$\log \dot M $\,[M$_\odot$\,yr$^{-1}$]} \\
        & dyn.  & spec.  & dyn. & spec. & spec. &evol. tracks\\
\noalign{\smallskip}
\hline
\noalign{\smallskip}
Tc\,1     & 0.59 & 0.81 & 32 & 34 & $-7.46$ & $-7.91$ \\
He\,2-108 & 0.57 & 0.63 & 32 & 34 & $-6.85$ & $-8.16$ \\
IC\,418   & 0.61 & 0.92 & 37 & 36 & $-7.43$ & $-7.82$ \\
NGC\,3242 & 0.61 & 0.63 & 79 & 75 & $-8.08$ & $-7.86$ \\
\noalign{\smallskip}
\hline
\end{tabular}
\end{flushleft}
\end{table}

\subsection{The mass distributions}

   \begin{figure}
   \includegraphics[width=88mm]{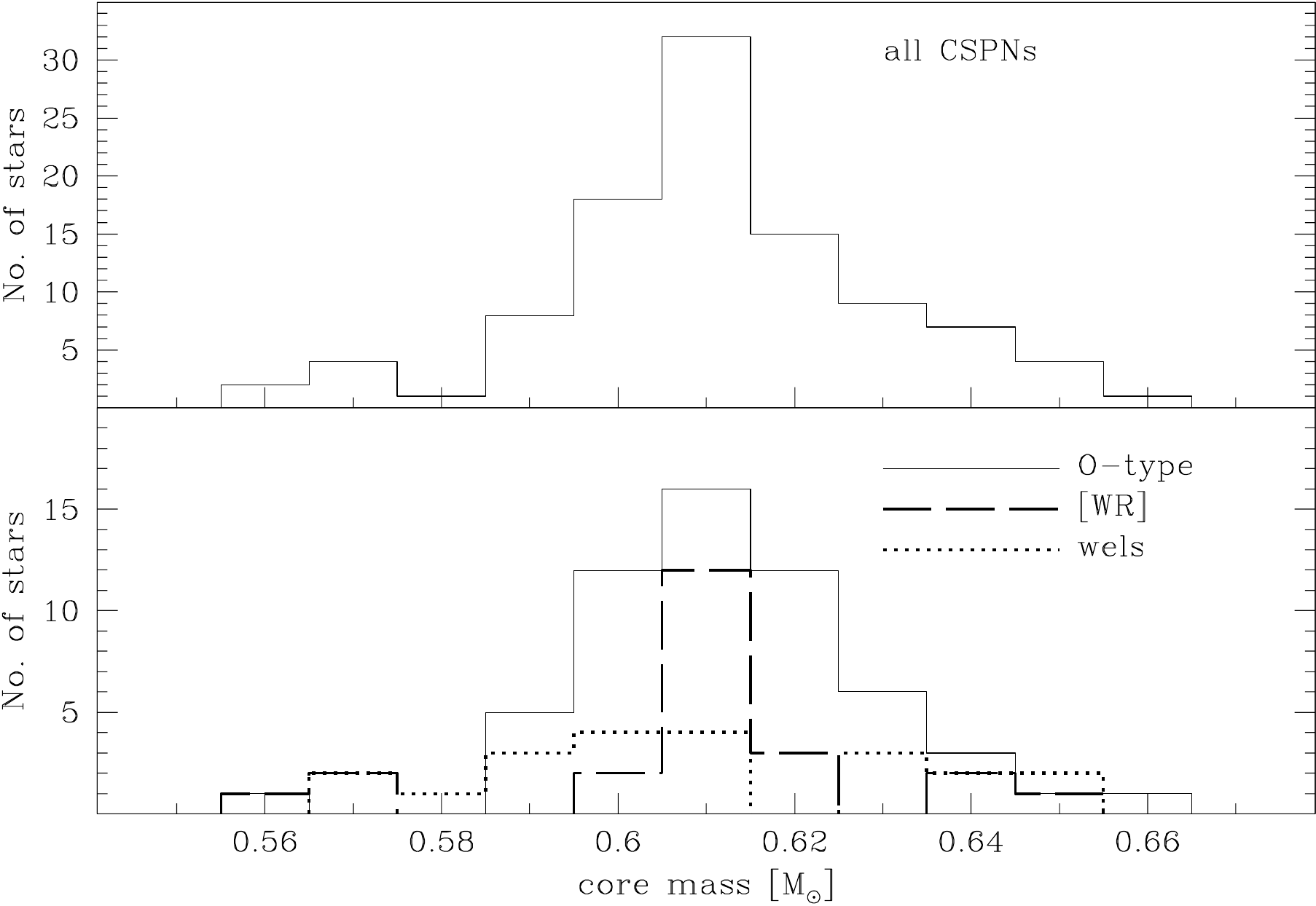}
   
     \caption{The CSPN mass histograms. Upper panel: the histogram of
     all modelled PNe. Lower panel: the histogram of different subgroups
     of the 101 PNe. The dashed line indicates [WR] stars, the dotted
     line {\it wels} and the solid line non-emission-line stars.}

         \label{cm}
    \end{figure}

In Fig.\ref{cm} the upper panel presents the mass distribution of our
whole sample of 101 PNe. All CSPNe masses fall into a narrow range,
$0.55 - 0.66\,M_{\odot}$, with a mean mass of $0.61\,M_{\odot}$. The
range of masses is almost identical to that of Tylenda et al. (1991) but
they obtained a smaller mean mass of $0.593\,M_{\odot}$ and their
distribution peaks at $0.58\,M_{\odot}$.

The lower panel of Fig.\ref{cm} presents masses  for the same types of
CSPNe as shown in Fig.\,\ref{hr}.  The non-emission-line stars show a
Gaussian mass distribution.  The hydrogen-deficient emission-line stars
seem to consist of two populations: one sharply peaked, containing [WR]
stars, and the other showing a wider spread, composed of [WR] and {\it
wels}.  The sharp peak consists, with a single exception, of hot [WO]
stars only. 

The presented histograms seem to suggest that hot [WO] stars form
a different group from the combined cooler [WC] and {\it wels} CSPNe.

\section{Comparing CSPNe and WDs}

\subsection{The histograms}    

The comparable birth rates of PNe and WDs  suggests that most white
dwarfs  go through the PN phase (e.g. Liebert et al. 2005). The mass
distribution in both samples should therefore be similar.

\begin{figure}
   \includegraphics[width=88mm]{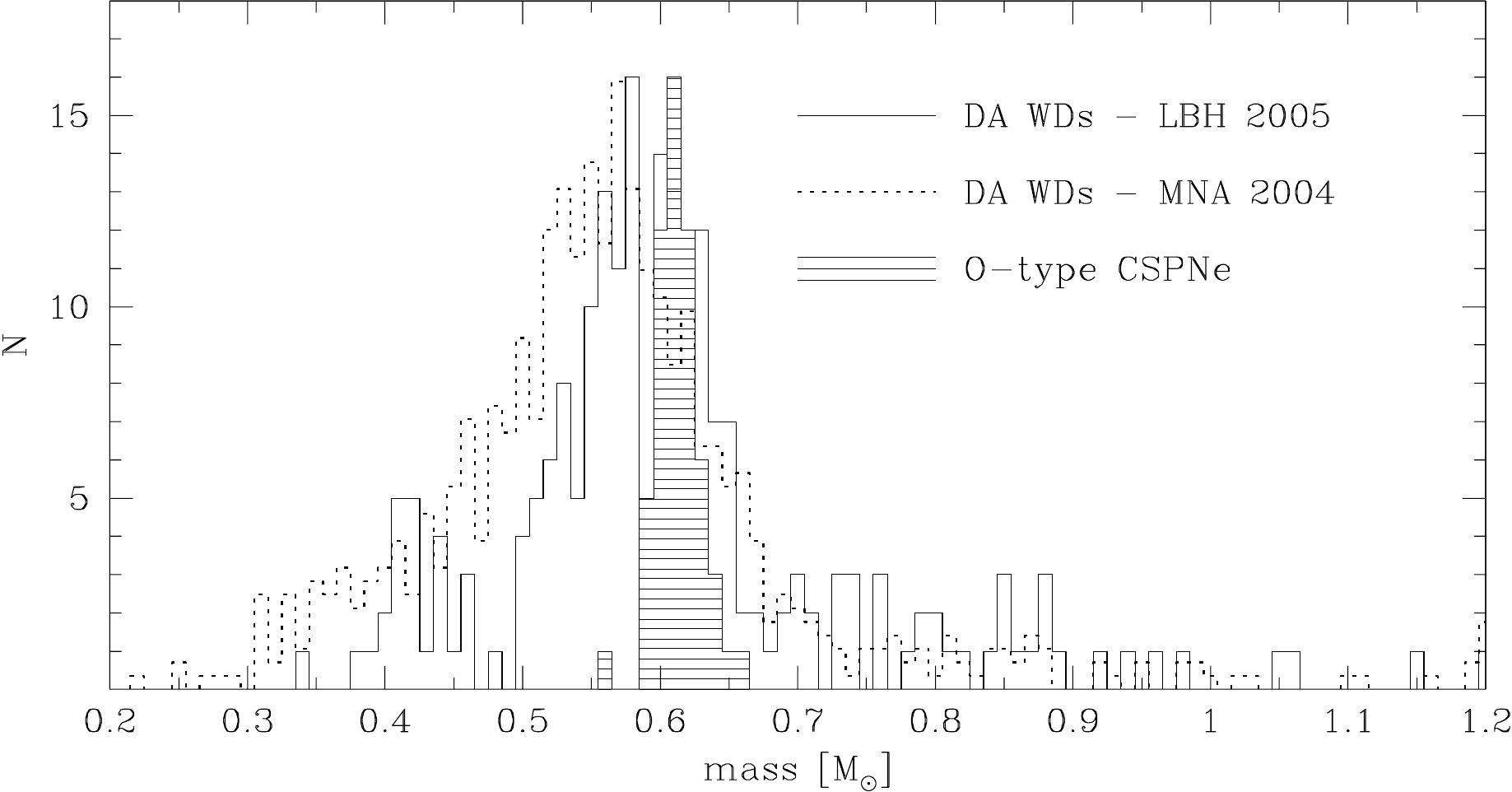}
   
     \caption{The mass distribution of non-emission-line O-type CSPNe 
     (shaded area) is compared to two DA white dwarf distributions
     of intermediate temperatures:
     thin line: data from Liebert et al. (2005); dotted line: data from
     Madej et al. (2004) which are more numerous, and are
     rescaled.}

         \label{hi}
    \end{figure}

Fig.\ref{hi} presents the histograms of our interpolated O-type CSPN
masses and the masses of DA white dwarfs from recent surveys. The WD
data of Madej et al. (2004), kindly provided by the authors, contain
1175 new DA WDs extracted from the Sloan Digital Sky Survey. The data of
Liebert et al.  (2005) taken from the electronic version of their
article, contain 347 DA WDs from the Palomar Green Survey. For
Fig.\ref{hi} we selected the objects with temperatures  between
15\,000\,K and 40\,000\,K. The two WDs histograms are not identical,
but both peak at similar values and show extended low- and high-mass
tails.  We plot the histograms using narrower bins than usually done for
WDs, optimized to the mass resolution of our CSPN data.  The difference
between the WD and CSPN distributions is striking.

First, the obtained CSPN masses are restricted to a much narrower range
of values than WDs, and are also much more sharply peaked. At face
value, this implies that only some of the WDs have gone through the PN
phase, in contrast to the conclusion from their similar birth rates
(Liebert et al. 2005).  Second, the two distributions peak at different
masses. Here a systematic  error cannot be excluded, as discussed
below.

\subsection{Hydrogen-rich vs. hydrogen-deficient}

Hansen \& Liebert (2003) point to a variety of WD mass distributions
with clear differences between hydrogen- and helium-rich cool
stars. Beauchamp et al. (1996) found for hot helium-atmosphere DB stars
a sharp peak lacking almost entirely of low- and high-mass components.
They also found that the DBA stars, which exhibit traces of atmospheric
hydrogen, show a distinctly different, broad and flat distribution.

The CSPN show an apparent difference between hydrogen-rich and
hydrogen-deficient mass distributions. The hydrogen-deficient stars show
a very narrow mass distribution; it is tempting to relate this to the
helium-rich DB and DBA populations. We use hydrogen-burning tracks to
derive these masses. The evolution after the thermal pulse leading to
helium burners is very complicated and not well understood (Werner \&\
Herwig 2006). This may not affect the derived masses too much: the
effect of a thermal pulse is to change the temperature of the star, but
as shown in Fig. 1, the isochrones have only a weak dependence on
temperature. The resulting offset in time (still very uncertain)
when accounted for can shift those CSPN masses towards higher values.

\section{Discussion}

\subsection{Uncertainties in mass determinations}

When comparing the CSPNe and WDs we have to remember that we compare
different spatial distributions. Because of their faintness the WD
observations are restricted to our nearest neighbourhood while PNe are
observed across the whole Galaxy. Nevertheless we didn't obtain
significantly different distributions for PNe at different distances.

Our mass determination relies on a single set of evolutionary tracks.
There are two possible sources of errors in the Bl\"ocker tracks. The
first is the early post-AGB evolution where the time scales depend on
how and when the AGB wind terminates. The Bl\"ocker tracks end this at
$T_{\rm eff} \sim 6000$\,K, (pulsation period of 50\,days) to agree with
the observations of detached shells around hotter stars but not around
cooler stars. A later termination would lead to an earlier start of the
ionization: in this case we would systematically overestimate the
masses. For a reduction of the post-AGB transition time by $10^3$\,yr,
the typical mass would reduce by 0.01\,M$_\odot$.

The second uncertainty is the mass-loss rate during the post-AGB phase.
For $M \sim 0.6\,\rm M_\odot$, the post-AGB mass-loss rate in the
Bl\"ocker models is 0.1 times the nuclear burning rate, but for
high-mass models the mass loss accelerates the evolution by 50\%\
(Bl\"ocker 1995). A higher post-AGB mass loss than assumed would reduce
our masses, but for the typical masses we find a very large increase
would be required. Table \ref{sp_mass} compares the Bl\"ocker mass-loss
rates with observed values, where we used the dynamical mass to
calculate the Bl\"ocker rate. For the three non-emission-line stars,
observed rates are higher by up to a factor of 3. This appears to be in
part related to the high luminosity derived by Kudritzki et al: if we
compare their rates with Bl\"ocker tracks at similar luminosity, then
the Bl\"ocker rates tend to be higher. The nuclear burning rate of $\dot
M_{\rm H} \sim -6.8$ exceeds the observed wind by a factor of four (more
for NGC\,3242). For this factor, the Bl\"ocker tracks would
underestimate the speed of evolution by only 10 per cent. We conclude
that the post-AGB mass-loss rates have little effect on the derived
masses. The exception is the {\it wels} star in the sample, where
the wind mass loss rate is comparable to the nuclear burning rate.

There is also an uncertainty in the dynamical age estimate. A later
acceleration would increase the ages by up to 50 per cent and shift the
mass peak from 0.61 to 0.60\,M$_\odot$. 

The WD mass determinations also suffer from simplifications and model
assumptions, in addition to the uncertainties concerning cool and hot WDs as
described in the Introduction.  One uncertainty is in contemporary plasma
physics, concerning the pressure broadening in a very high density 
plasma (Madej et al. 2004).  The mass-radius relations used depend on the
assumed mass of the hydrogen layer.  Napiwotzki et al. (1999) compared
estimates from different studies and concluded that the gravities obtained
from spectroscopic method suffer from systematic errors of up to 0.1\,dex in
$\log\,g$. This corresponds to an offset in masses of about 0.02\,M$_\odot$
and could, in principle, explain the difference in peak masses between WDs and
CSPNe. The width of the peak may also be narrower than derived from the
models. Nevertheless, the wide tails of the mass distribution are not in
doubt.

\subsection{Time scales, birth rates and binarity}

The derived CSPN mass distribution combines the effects of the birth
rate as function of mass, and the observable life time of the PN. The
latter depends on mass as indicated in Table \ref{time}. The period of
visibility is defined here as beginning when the star reaches $T_{\rm
eff}=25\,10^3$\,K, and ending either when the star enters the cooling
track (defines as $\log L = 3.00$) or when the age of the nebula is
$10^4$\,yr, whichever comes earlier.  Our histogram should be corrected
for the difference in visibility time. This increases the
number at high CSPN mass only by a factor of up to 10, and brings the high 
mass tail  in somewhat better agreement. We may also have a sample bias
against high masses, as these are not expected in the Bulge objects. The 
de-selection of bipolar objects may have removed a few higher-mass nebulae
in the disk.

CSPNe with $M<0.56\,\rm M_\odot$ would not produce a visible PN, as the
post-AGB transition time becomes too long ('lazy PNe'). In the sample of
Liebert et al. (2005), 30 per cent of white dwarfs have masses in this range,
and 50 per cent in the sample of Madej et al. (2004). However, the sharp
drop in the CSPN mass distribution below 0.60\,M$_\odot$ occurs at too high
mass to be affected.

\begin{table}
\caption[]{\label{time} Bl\"ocker track time scales: PN visibility is
defined as between $\log T_{\rm eff} = 4.4$ and either a
nebular age $t = 10^4\,\rm yr$ or a stellar luminosity $\log L = 3.0$,
whichever occurs earlier
  }
\begin{flushleft}
\begin{tabular}{llllllllll}
\hline
  Mass  [M$_\odot$]  & $t_{\rm start}$\,[yr]  & $t_{\rm end}$\,[yr] &
 $t_{\rm visibility}$\,[yr] \\
\hline
0.546 & $90\,10^3$  & - & - \\
0.565 & $4\,10^3$   & $10\,10^3$   & $6\,10^3$\\ 
0.605 & $1.5\,10^3$ & $7.4 \, 10^3$ & $5.9\,10^3$\\ 
0.625 & 660         & $3.6\, 10^3$  &  $2.9\,10^3$\\ 
0.696 & 100         & 880           & 780 \\
0.836 & 100         & 840           & 740 \\
0.940 & 12          & 90            & 78 \\
\hline
\end{tabular}
\end{flushleft}
\end{table}

Hansen \& Liebert (2003) argue that both the high- and low-mass tails in
WDs distribution can be a result of  binary evolution. Merging leads to
high-mass WDs while a close companion stripping the envelope can cause
an early termination of the evolution and produce a low-mass helium WD.
Both channels together may account for some 10 per cent of all WDs (Moe
\&\ de Marco 2006). Therefore the histogram for single WDs could be
narrower. Close binary evolution can affect the PN phase as well,
leading to strongly non-spherical nebulae. Our model analysis assumes
spherical symmetry, and we did not analyze bipolar nebulae. Our
selection therefore favours single CSPNe and rejects low-mass CSPNe
in interacting binaries. Thus, the CSPN histogram (Fig.\,\ref{hi}) is
biased toward single-star evolution, while the WD histogram includes
binary broadening. This may affect the tails of the WD histogram but is
not expected to affect the main peak.

Moe \&\ de Marco (2006) predict a number of PNe in the Galaxy of around
46000. Based on local column densities, Zijlstra \&\ Pottasch (1991)
derive an actual number of 23000, suggesting that only about half the stars
which could produce a PN, do so. This comparison is limited by our
knowledge on the time a PN remains observable.  Moe \&\ de Marco (2006)
predict a birth rate of PNe of $1.1 \times 10^{-12}\,\rm PNe\, yr^{-1}\,
pc^{-3}$, comparable to the current, local WD birth rate of $1.0 \times
10^{-12}\,\rm PNe\, yr^{-1}\, pc^{-3}$. Again assuming only half their
predicted number of PNe is actually observed, the expectation is that
half of all WDs have passed through the PN phase.

\section{Conclusions}

We show that the mass distribution of CSPNe is sharply peaked at
$M=0.61\,\rm M_\odot$. The published WD mass distributions show a much
broader distribution peaking at a lower mass of $M=0.59\,\rm M_\odot$.
Part of the difference in the peak may indicate faster evolution during
the early post-AGB phase than assumed in the Bl\"ocker tracks.  CSPN
mass-loss rates cannot explain the difference. However considering
the uncertainty of 0.02\,M$_\odot$ in the WD mass estimations both peaks
are in reasonable agreement. About 30 per cent of WDs have too low
masses to have passed through the PN phase.

\begin{acknowledgements}
We thank our referee Ralf Napiwotzki for important comments.
This project was financially supported by the ``Polish State Committee for
Scientific Research'' through the grant No. 2.P03D.002.025 and by a NATO
collaborative program grant No. PST.CLG.979726. AAZ and KG gratefully
acknowledge hospitality from the SAAO.
\end{acknowledgements}

\end{document}